\documentclass[12pt,twoside]{article}
\usepackage{graphicx}
\usepackage{epsf}
\usepackage{epsfig}
\begin{document}
\newcommand{\be}{\begin{equation}}
\newcommand{\ee}{\end{equation}}
\newcommand{\bea}{\begin{eqnarray}}
\newcommand{\eea}{\end{eqnarray}}

\title{
{\bf Thermodynamics - a valuable approach to multifragmentation? } }
\author{  Regina Nebauer$^{*+}$ and J\"org Aichelin$^+$ \\
  $^+$SUBATECH \\
  Universit\'e de Nantes, EMN, IN2P3/CNRS \\
  4, Rue Alfred Kastler, 44070 Nantes Cedex 03, France \\
  $^*$ Universit\"at Rostock, Germany} 
\maketitle
 
\begin{abstract}
Since years it has been vividly debated whether multifragmentation is a thermal 
or a
dynamical process. Recently it has been claimed \cite{toek1,po} that new data
allow to decide this question. The conclusion, drawn in these papers, are,
however, opposite. Whereas \cite{toek1} states that the behavior of different
observables as a function of the fragment multiplicity excludes a thermal
origin of the fragments in \cite{po} it has been argued that data show 
a first order phase transition between a liquid and a gaseous phase. 
It is the aim of this paper to show that both conclusions are premature. They
are based on the salient assumption, that the system is sufficiently large to be
susceptible to a canonical description. We will show that this is not the case.
A micro canonical approach describes the data as good as dynamical 
calculations. Hence the quest for the physical origin of multifragmentation 
continues.
\end{abstract}

\section{Introduction}
Since almost a decade the study of multifragmentation characterized by the
multiple production of intermediate mass fragments ($3 \le Z \le 20$)
is one of the central
issues in intermediate energy heavy ion collision studies. For beam energies in 
between 50A~MeV and 400A~MeV multifragmentation has been identified as
the dominant reaction channel and up
to 15 intermediate mass fragments (IMF's, $~Z\ge~3$) have been observed in a single
event. 

The mechanism of multifragmentation, however, remained rather debated because
the different mechanisms proposed predict the same
functional dependence for several key observables. If the  
disintegration of the nucleus is instantaneous each nucleon keeps its
momentum and one expects 
an average fragment kinetic energy of $3/5 E_{F}$  \cite{gol} independent of the
fragment size , 
where $E_{F}$ is the Fermi energy.
The same independence one expects if the fragments are formed very late,
after the system has been expanded while maintaining thermal equilibrium.
This requires that the disintegration is very slow.
Here the average kinetic energy is 3/2 T, where T is the
temperature at freeze out. The same ambiguity one finds for the mass yield
where thermal\cite{fi} and non thermal systems \cite{hu,hue} show the same form. 
Hence more complicated observables have to be employed to distinguish between
the different possible reaction mechanisms. 

Recently it has been conjectured
by T\~oke et al. \cite{toek1} that one can distinguish between a statistical
process and a dynamical process by investigating several observables as a 
function of the fragment multiplicity. In analyzing their data, obtained in 
their limited acceptance region, they made the following observations:
\begin{enumerate}
\item The average transverse kinetic energy of the intermediate mass fragments 
(IMF's) is independent of the number of observed fragments.
\item The average light charged particle (LCP) multiplicity as well as 
the total kinetic
energy of the LCP's is also independent of the number of observed
fragments.
\end{enumerate}
They considered this observation as contradictory to the assumption that
the system is in equilibrium and argue as follows: Independent of the unknown
impact parameter b the energy per nucleon and hence the temperature
in the center of mass is almost identical. 
By varying b one changes the size of the source but not
its temperature. Indeed, the observed average kinetic energy of the LCP's
is independent of the number of IMF's. If the temperature of the system
is constant in thermal equilibrium the ratio between LCP's and
IMF's is determined by the chemical potential and hence fixed.
Thus the increase of the particle  number 
(sum of free nucleons and those entrained in fragments) in the observed phase space interval 
has to be shared between the fragments and the free nucleons.
Due to 2.) this is obviously not the case.

The opposite conclusion has been drawn from the analysis of experimental data by
\cite{po}. In this paper it is the isotopic yield ratio of fragments which 
is used to 
determine the temperature of the systems. Plotted as a function of the
excitation energy of the system the temperature shows a plateau. The claim that
the system is in thermal equilibrium is based on the similarity
of this observation with the latent heat observed in a first order phase 
transition in infinite systems.

Both interpretations of the experimental data rely on the assumption that the
equilibrated source is sufficiently large to be treated as a canonical system.
It is the first purpose of this article to show that this salient assumption 
is not justified. Using a microcanonical statistical model \cite{bon95}
we reproduce the results of ref. \cite{toek1}. This renders their 
conclusions premature. Taking the temperature fluctuations in finite size
systems seriously we will show that the variance of the temperature 
distribution is not negligible and hence one cannot distinguish between 
a constant and an increasing temperature as a function of the excitation energy
of the system. Furthermore
the temperature fluctuations give rise to a  nontrivial relation between
the apparent temperature measured by the fragment yields and the temperature
of the emitting source. Thus the conclusions of ref.\cite{po} that data show an
allusion of a first order phase transition are premature as
well. The experimental determination of the mechanism which causes
multifragmentation remains to be a challenge. 

The second purpose of this article is to show that microcanonical statistical
models predict strong correlations for systems as small as those observed in
heavy ion reactions. 

The third purpose is to show that - even worse - the observables
discussed in ref. \cite{toek1} do not allow to distinguish between 
a dynamical  and a statistical reaction scenario. 
For that purpose we compare the results of the microcanonical statistical model
\cite{bon95} with those obtained by Quantum Molecular Dynamics 
simulations \cite{aic}. The
analysis of the latter shows a nonstatistical origin of the
fragments\cite{neb,reg,qmdh}
(although the model itself is able to follow the evolution of a system
towards equilibration if there were any).

\section{Microcanonical and Canonical Systems}

Before we start out with the detailed calculation a reminder on the 
differences between the microcanonical and the canonical
approach to study systems in equilibrium is appropriate. For this discussion we will be 
guided by the question at hand: what can we learn about the system by analyzing
observables measured in a subsystem. This subsystem may be the limited part 
of the phase space which is free from a contamination by preequilibrium 
emission or may be the ensemble of fragments in experiments
where nucleons cannot be measured. This question can
be discussed best if we divide the system,
which is characterized by its macroscopic parameters, the energy E, 
the particle number N and the volume V,
into two parts, the observable and the unobservable part of phase
space. 

A microcanonical description of an equilibrated system is based on the assumption that
in statistical equilibrium each microstate which is compatible with (E,V,N) 
is occupied with the same probability. $g_{obs}
(E_{obs},V_{obs},N_{obs})$ is the number of microstates available in the
observable sector which is characterized by $(E_{obs},V_{obs},N_{obs})$. 
The number of microstates of the whole system is consequently
\begin{equation}
G(E,V,N)=\sum_{N_{obs}}\sum_{E_{obs}} 
g_{obs}(E_{obs},V_{obs},N_{obs}) 
\cdot g_{unobs}(E-E_{obs},V-V_{obs},N-N_{obs}).
\end{equation}
$V_{obs}$ is considered as constant here.
The entropy of the system is given by $S^{mc}= 
k \ log \ G(E,V,N)$ (k being the Boltzmann constant) whereas the entropy of
the observable subsystem is 
$S_{obs}^{mc}= k \ log \ g_{obs} (E_{obs},V_{obs},N_{obs})$.
Knowing the functional dependence of S on E and N we can easily calculate 
the temperature $1/T =
{\partial S\over \partial E}$ and the chemical potential ${\mu \over T} =  {\partial S\over
\partial N}$.
For the combined system with
\be 
 G(N,E,E_{obs}) = \sum_{N_{obs}} g_{obs}(N_{obs},E_{obs})
 g_{unobs}(N-N_{obs},E-E_{obs})
\ee  
one finds  
\bea
{k\partial ~ ln G \over \partial E_{obs}} &=& \sum_{N_{obs}}  
{k\partial ~ ln g_{obs}(N_{obs},E_{obs}) \over \partial E_{obs}} +
{k\partial ~ ln g_{unobs}(N-N_{obs},E-E_{obs}) \over \partial E_{obs}}\\
&=&\sum_{N_{obs}}  
{k~\partial ~ ln g_{obs}(N_{obs},E_{obs}) \over \partial E_{obs}} -
{k~\partial ~ ln g_{unobs}(N-N_{obs},E_{unobs}) \over \partial E_{unobs}}\\
&=&\sum_{N_{obs}}  
{1\over T_{obs}} -{1\over T_{unobs}}
\eea
For fixed values of E,N we may find in the both subsystems different values
$E_{obs}$'s and $N_{obs}$'s and consequently a distribution of $T_{obs}$'s and
$\mu_{obs}$'s.  Only at the maximum $
{k\partial ln G(N,E,E{obs}) \over \partial E_{obs}} = 0$ the temperatures in
both subsystems agree. 
Therefore generally it is not possible to infer from the 
temperature, chemical potential, particle number  or  energy of the 
observable subsystem the corresponding values of the whole system or the other
subsystem and vice versa. 

This is of course completely counterintuitive because from daily life we are
used to large systems. Because for a given energy of the total system the fluctuations
of the temperature in a subsystem are $\propto {1\over \sqrt{N}}$, where N is the
number of particles in this subsystem, in macroscopic systems the fluctuations
of the temperature can be neglected. In other words, if the system is 
sufficiently large $(E \rightarrow \infty ,
N \rightarrow \infty) \quad g_{obs}\cdot g_{unobs}$ is very sharply peaked 
and the sums in eq. 1 can be replaced for all practical purposes by the 
largest term. Then the entropy becomes additive 
and the temperatures and chemical potentials in the subsystems are identical.
Under this condition the microcanonical 
and the canonical calculations coincide. If this is not the case 
the (isolated) system is not susceptible to a canonical but only to a
microcanonical treatment.

For the interpretation presented in \cite{toek1,po} it is crucial 
that the system is large and energetic enough to justify a canonical treatment. 
In ref. \cite{toek1} the observed subsystem can be identified with the limited
acceptance region of the experiment, in ref.\cite{po} with the subsystem 
of fragments which is used to determine the apparent
temperature.  It is this apparent temperature which has a plateau as a 
function of the excitation energy of the equilibrated source.

To verify if the nuclear system is sufficiently large to justify a canonical
description we employ one of the state of the art statistical models, 
the Statistical Multifragmentation Model (SMM), which has been developed by the Copenhagen group
and has later been improved by  Botvina \cite{bon95}.
There all possible microstates of the nuclear system are carefully elaborated. 
This model has been employed frequently to interpret heavy ion
reaction data. It suffers, however, as every model of this kind, 
from 3 unsolved problems: a) from the problem how to treat
particle unstable nuclear levels, b) from the not well known level density at
large excitation energies and c) from the fact that there is
no experimental information on the freeze out volume.
The results of statistical model calculations depend on the proposed solution
of these problems \cite{gs}. Therefore, to study the conjecture of \cite{po},
where this problem becomes crucial, we employ in addition
a simpler microcanonical model which allows for an analytical solution. 
It serves
as well the purpose to understand qualitatively the fluctuations observed in the
SMM calculations. For a general introduction to the statistical physics of 
multifragmentation we refer to ref.\cite{gr}.

\section{Comparison with data}

For our studies we use data and simulations for 
central collisions of the reaction 50A~MeV Xe~+~Sn where precise data are
available, taken by the INDRA collaboration\cite{sal,mar}. Both models, SMM and QMD, have 
been extensively used to interpret this reaction \cite{reg,sal}. As said the analysis of the QMD calculations shows that in this approach
the system never passes through a state of thermal equilibrium \cite{neb,qmdh}
whereas the application of SMM is only justified if such a state is formed.
Thus the reaction scenarios in these models are orthogonal.
For the comparison with the data both calculations have been filtered with the
experimental acceptance. To model the centrality it turns out \cite{reg} that
one has to require for the (filtered) QMD simulations that the total transverse
energy of LCP's is larger than 450 MeV. For the statistical model 
calculations we use a slightly different centrality cut.
For a comparison of both cuts, whose difference is of no importance here,
we refer to ref. \cite{reg}. The increase of the average transverse energy of
the fragments as a function of their mass, observed in the experiment, 
is larger than predicted in SMM \cite{sal} calculations. 
Therefore one has to modify the statistical model calculations by adding a fourth
system parameter,  a collective energy, which is parameterized as
$ E_{coll}=c*A $,
where c is a parameter which remains to be determined and A is the fragment
mass. The best agreement between experiment and SMM calculation is obtained
with the following set of input parameter:  
\begin{tabbing}\hspace*{4.5cm}\=\kill
freeze out density:\>$1/3 \rho_0$\\
source size:\> Z$_S$=78 \ A$_S$ = 186\\
excitation energy:\> E$_{thermal}$=7A~MeV \ \  E$_{coll}$=2.2A~MeV\\
\end{tabbing}

Even central collisions at intermediate energy 
have a binary character \cite{reg} and consequently emission of particles from
residues close to beam or target velocity spoils the spectra of particle
emission from a possible thermal 
source at rest in the center of mass. 
Therefore a meaningful comparison between statistical model calculations and
experimental data is only possible around $\theta_{CM} = 90^o$.
We subdivide the experimental data and the QMD simulations into 
two equal size $2 \pi$ intervals:\\
\centerline{$B_{obs}$ : \ \ $60^o\le\theta_{CM}\le120^o$}
\centerline{$B_{unobs}$ : \ \ $\theta_{CM} < 60^o,\theta_{CM} > 120^o$.}
In $B_{obs}$ we observe a flat angular distribution and a constant average
energy of IMF's and LCP's \cite{mar} as a function of the emission angle, both
being prerequisites for a statistical equilibrium. In $B_{unobs}$, 
on the contrary,in the thermals language
a pre\-equilibrium component is superimposed to the thermal component.
QMD which does not separate the phase space into two contributions but gives an
continuous distribution for all observation angles describes this region well 
\cite{reg}.

\begin{figure}[h]
\vspace{-1.5cm}
\epsfxsize=12.cm
$$
\epsfbox{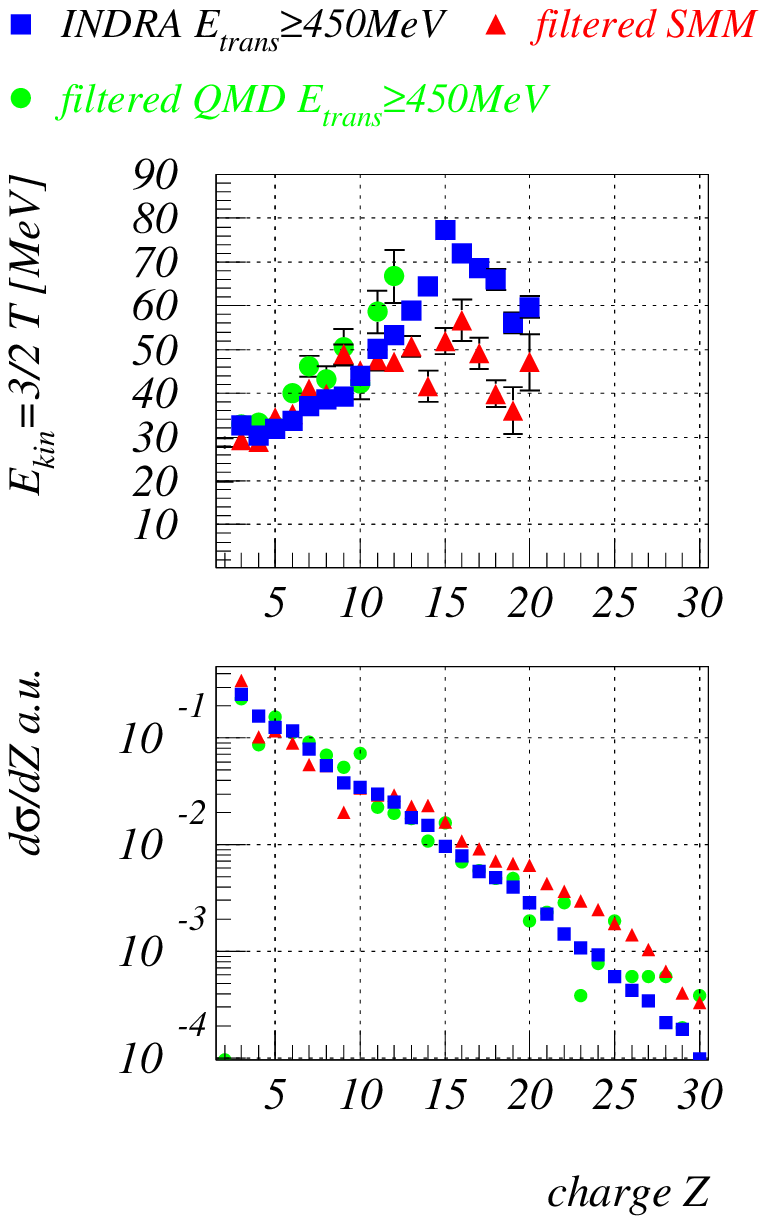}
$$
\vspace{-1cm}
\caption{\textit{Fitted slopes and the charge distribution for QMD, INDRA
and SMM data in $60^o\le\theta_{CM}\le120^o$ for 50A~MeV Xe + Sn, central
collisions.}}
\label{temz}
\end{figure}

The theoretical as well as the experimental fragment energy distributions
show an exponential shape. In fig.~\ref{temz} we
display on top the slope of the kinetic energy spectra of the fragments. 
The inverse slopes T are converted into
an energy. On the bottom we show the charge yield distribution. We display the
results for QMD and SMM calculations in comparison with the INDRA data. As one
can see, these are well reproduced in both theories, underlining the above
mentioned observation that these observables are not sensitive to the 
reaction mechanism. 

\begin{figure}[hp]
\vspace{-3cm}
\epsfxsize=12.cm
$$
\epsfbox{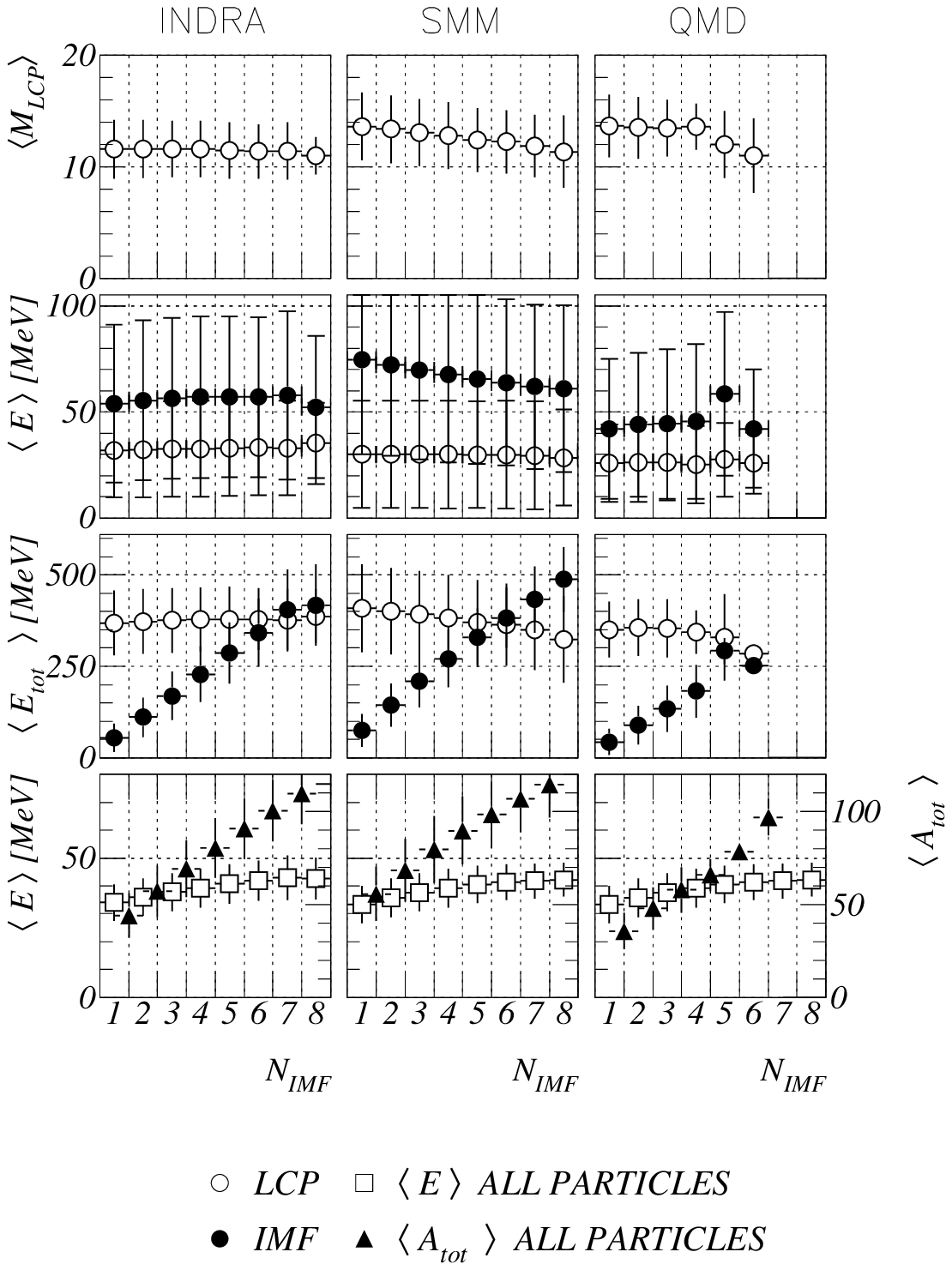}
$$
\vspace{-2cm}
\caption{\textit{As a function of the IMF multiplicity observed in 
$60^o\le\theta_{CM}\le120^o$ we display for central events 50A~MeV Xe + Sn
several observables in $60^o\le\theta_{CM}\le120^o$:
The LCP multiplicity (top row), the average kinetic energy of
IMF's and LCP's (second row), the total energy of IMF's and LCP's (third row),
the average kinetic energy of all particles and the number
of nucleons(forth row).}} 
\label{imf1}
\end{figure}

In figure~\ref{imf1} we plot for $B_{obs}$ different observables
as a function of the IMF multiplicity. The
left column displays the experimental results for the central reaction 
50 AMeV Xe + Sn as measured by the INDRA collaboration. 

In the first row we display the LCP multiplicity as a function of the number 
of observed IMF's. At higher energies, where the number of LCP's is much larger, 
the  LCP multiplicity
is frequently considered as a measure for the centrality of the reaction (what
is confirmed by the QMD calculations). As can be seen from the panel,
in our case the multiplicity of LCP's is independent of the IMF multiplicity.
Thus one may conjecture that  the
(central, $E_{trans}\geq450~MeV$) events
with different IMF multiplicities in $B_{obs}$ have the same average impact
parameter. Also the average kinetic energy of LCP's and IMF's, displayed
in the second row, is independent of the number of IMF's. 
The third row displays the sum of the kinetic energies  of the IMF's and LCP's,
respectively. We observe, as expected from row 1 and 2, a constant value for the LCP's 
and a linear increase for the IMF's.
The fourth row displays the total kinetic energy in $B_{obs}$ divided by the 
sum of the number of LCP's and IMF's. In a canonical system of noninteracting 
particles this quantity is related to the temperature by $T = {2\over 3}{E\over
N}$. We see that the average value decreases with decreasing
fragment number and the fluctuation are considerable (
$<\Delta E/\overline{E}> = \sqrt{ \overline{E^2} - 
{\overline{E}^2}}/\overline{E}~=~0.2)$. 
The fourth row shows as well the total number of nucleons (free or as part of
the fragments) in $B_{obs}$. It varies by almost a factor of 3
although the number of LCP's stays constant.

These observations, although presented here for another reaction, agree well
with the findings of ref. \cite{toek1}. Thus, 
if their arguments were valid, we would arrive at the same conclusion.

However, we are dealing with a rather small system. Fluctuations 
between $B_{obs}$ and $B_{unobs}$ may be important and hence the system may be
too small for a canonical description. To see whether this is true we 
performed the same analysis using the microcanonical statistical model.
The results are presented in the second column 
of fig. 2. We see that the statistical model results agree well 
with experiment. The results are, however, quite different from those of a
canonical description of the data. There $\mu$ and T are identical in both
subsystems once E,N,V are given and so is the ratio
between IMF's and LCP's which is  fixed by the chemical potential. Thus fluctuations
(between the two small subsystems $B_{obs}$ and $B_{unobs}$) and correlations
(due to the conservation of N and E) are not only important but essential for 
the results.

Does this result mean that data prove that multifragmentation 
is a statistical process?
To answer this question we performed calculation with QMD \cite{aic} The results are presented in the third column. Besides of the too
small average fragment energy ( which is a consequence of a shortcoming of
QMD, the artificial long range of the attractive nuclear potential which 
suppresses the Coulomb repulsion \cite{aic} and therefore decreases the average 
fragment energy) also the results of QMD agree in the error bars with those of
experiment.

Thus we arrive at the conclusion that even this quite involved analysis of
the fragment production does not allow to distinguish between a statistical and a
dynamical origin. One should keep in mind, however, that QMD is able to 
describe both, $B_{obs}$ and $B_{unobs}$, whereas a comparison with 
SMM is limited to $B_{obs}$ because in $B_{unobs}$ nonstatistical processes
dominate the fragment production.

\begin{figure}[h]
\vspace{-2.5cm}
\epsfxsize=12.cm
$$
\epsfbox{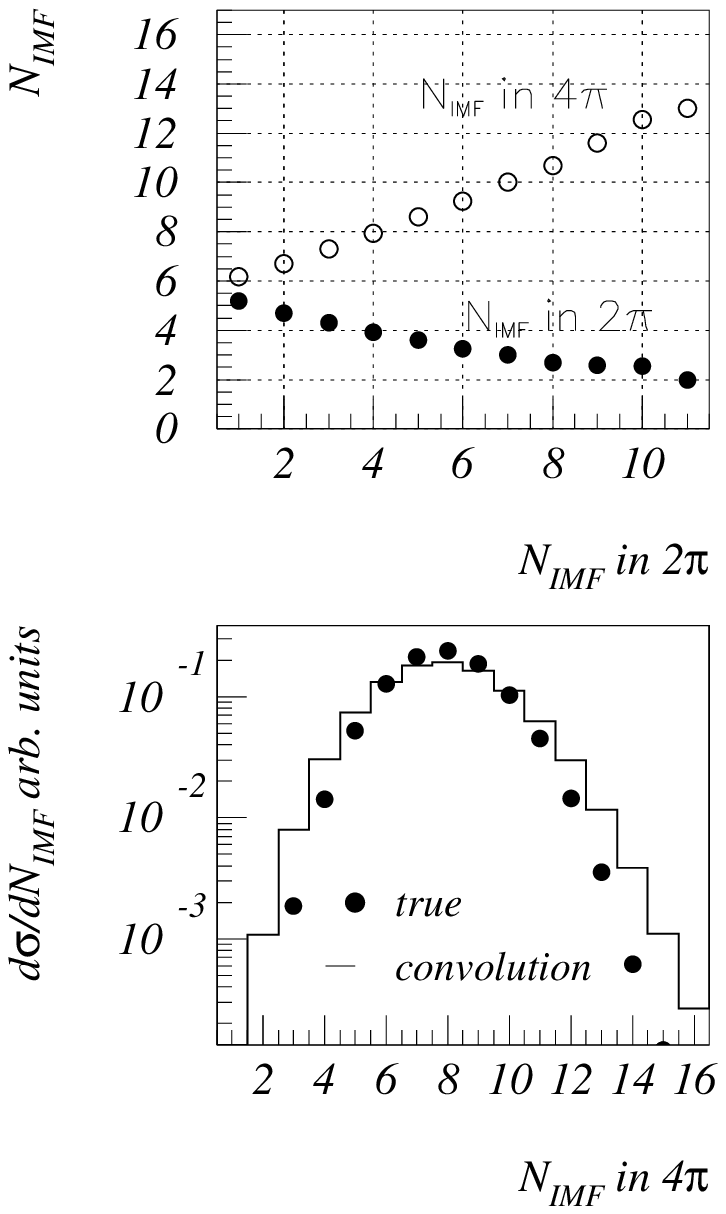}
$$
\vspace{-1cm}
\caption{\textit{Correlations as predicted by SMM calculation.
We display the mean value of IMF's in the $B_{unobs}$ and in $4\pi$ as a 
function of the fragment multiplicity in  $B_{obs}$ (top). Top right we display 
the actual IMF multiplicity
distribution as compared to a convolution of the distributions observed in
$B_{obs}$ and $B_{unobs}$. The statistical model calculations are done for central 
reactions 50A~MeV Xe + Sn}}
\label{imf2}
\end{figure}

How the correlations in the statistical model calculation 
show up in detail is shown in  fig.~\ref{imf2} where we display 
the SMM results before filtering.
The upper panel shows the fragment multiplicity  in $4\pi$ and in $B_{unobs}$
as a function of the fragment multiplicity observed in $B_{obs}$. 
Of course, without filtering $B_{obs}$ and $B_{unobs}$ are two arbitrary
$2\pi$ bins which should not differ. However,
we observe strong correlations between the fragment multiplicities in the 
two $2\pi$ intervals. In the lower panel we
compare the true fragment multiplicity distribution in $4\pi$ 
with that obtained by a 
convolution of the multiplicity distribution observed in one $2\pi$
subsystem. If no correlations were present we would expect for $4\pi$ 
a convolution of the distribution observed in $2\pi$. This is obviously
not the case and consequently it is impossible 
to infer in systems as small as this 
the  multiplicity distribution in $4\pi$ from a $2\pi$ subsystem.

\begin{figure}[h]
\vspace{-2.cm}
$$
\epsfxsize=10.cm
\epsfbox{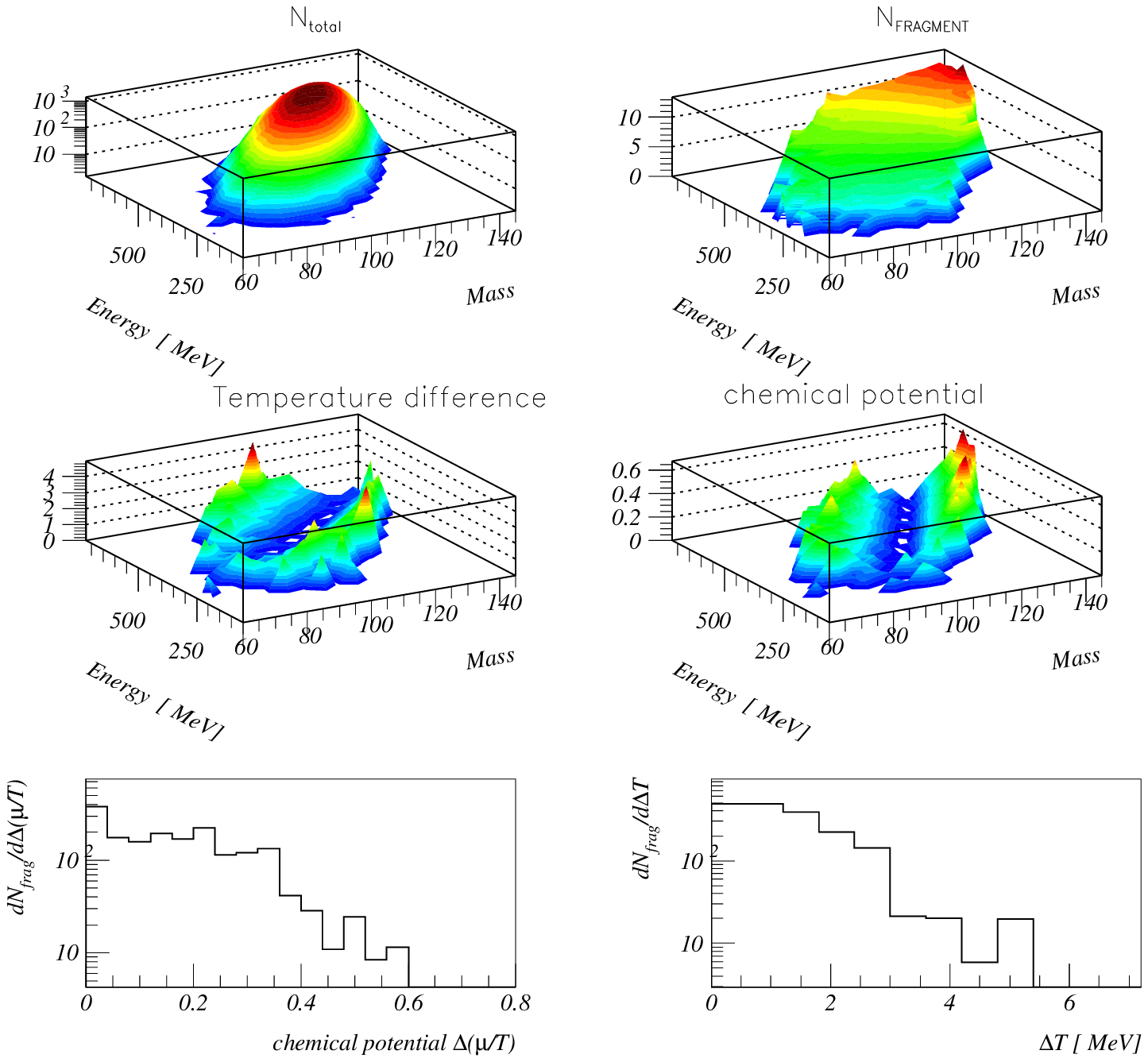}
\vspace{-1cm}
$$
\caption{\textit{Top: Number of events and number of fragments as a function
of the total mass of all fragments and of their total energy. Middle:
Distribution of the temperature difference and the difference of the chemical
potential. Bottom: Distribution of the absolute value of the temperature
difference
distribution and that of the distribution of the difference of the 
chemical potentials between the subsystems formed by fragments and 
LPC's, respectively, for a SMM calculation adjusted to describe the INDRA 
results}}
\label{et}
\end{figure}

\section{Temperature fluctuations in a small system}

To address the conjecture of the second paper \cite{po} it is important to
now the temperature fluctuations in the relevant subsystem  under the 
condition that the total energy of the system is constant. This question is not
easy to address because the relevant subsystem is the environment of
those fragments which are used to determine the temperature by calculating
isotope ratios. This information is not available in SMM. Therefore we assume
that the temperature fluctuations of the subsystem consisting of the
intermediate mass fragments $Z\ge 3$ is a good measure for those of the 
above mentioned observable. We use the above mentioned microcanonical SMM
calculations and identify the two subsystems with the
fragments with $Z\ge 3$ and the LPC's, respectively. To calculate the
temperature fluctuation we employ eq.3.
 
In fig. \ref{et}, top,  we display the number of events (left) 
and the average number of IMF's (right) as a function of the
total number of nucleons entrained in the fragments and the total fragment
energy. We see a rather broad distribution. In the middle we present the
difference of the temperatures and of the chemical potential in the
two subsystems (LPC's and IMF's) as a function of the total mass and the 
total energy of all IMF's. For calculating the temperature difference we have 
assumed that 
${1\over T_1} -{1\over T_2} \approx {\Delta T\over {T^2 + \Delta T^2/4}}$ 
where T is assumed to be  7 MeV. We see a rather broad distribution of the
chemical potentials and of the temperatures. The probability distribution 
of fragments to be emitted in a microstate which shows a 
$\mid \Delta T \mid$ and a chemical potential difference of  $\mid \Delta \mu
\mid$, respectively, 
between the two subsystems
is plotted in the bottom row. The chemical potential $\mid \Delta \mu \mid$
in units of the temperature fluctuates by about to 40\% around the mean value 0
whereas the variation of the distribution of the temperature difference 
between the subsystems is about 1.7 MeV. 
More precisely the temperature in the subsystem of the fragments
cannot be determined. 

To show that these results are generic and do not depend decisively on the
mentioned problems of these microcanonical statistical model approaches
we employ a much 
simpler model to confirm the order of magnitude of the above result.
It has the advantage that is allows for analytical results. 
Although this model neglects the long range Coulomb force
which make thermodynamics much more complicated it is sufficiently realistic to
understand the physical origin of the fluctuations.

We consider a noninteracting system consisting of $N_F$ fragments and $N_N$ free
nucleons. The energy $E = E_F + E_N$ of the system is constant. We calculate
the fluctuation of the total fragment energy and hence the fluctuation
of the temperature if only fragments are measured. The temperature of the whole
system is fixed once density , particle number and total energy are given.

 The probability to have a total fragment energy $E_F$  
is proportional to the available number of microstates
for this division of the energy: 
\be
P(E_F,\rho,N_F) = {e^{S(E_F,\rho,N_F)/k +S(E-E_F,\rho,N_N)/k} \over
e^{S(E,\rho,N_F+N_N)/k}}.
\ee
Assuming that we can use the ideal gas entropy
\be
S(E,\rho,N) = Nk({5\over 2} + ln ({\rho\over h^3}({4\pi m E \over 3N})^
{3\over 2}
))
\ee
the probability to find an energy fraction of $E_F/E$ for the
total energy of the fragments is given by
\be
P(x = E_F/E,N_F,N_N) 
= {\Gamma(2 + {3N\over 2})\over\Gamma(1 + {3N_F\over 2})
\Gamma(1 + {3N_N\over 2})} (1-x)^{3N_N/2}x^{3N_F/2} 
\ee
where N = $N_F + N_N$.
For the standard deviation of the energy 
$\Delta E_F = \sqrt{ \overline{E_F^2} - 
\left.\bar {E_F}\right.^2}$  we find
\be
{\Delta E_F \over E} = \sqrt{{3N_F+2 \over 3N+4} {6(N-N_F)+4\over 
(3N+6)(3N+4)}} \quad ; \quad
{\Delta E_F \over \overline{E}_F} = {\Delta T_{F} \over T_{F}} = 
\sqrt{{6(N-N_F)+4\over (3N_F+2)(3N+6)}}.
\ee
For central reactions 50A MeV Xe+Sn we have typically $N_N = 50$ and $N_F= 8$.
Thus we find 
${\Delta T_F \over \overline{T}_F} = 0.24$ and hence about the same value as in
the SMM approach. Thus the order of magnitude of the fluctuation is given by the
size of the system only and does not depend on the details of the Hamiltonian or
the freeze out volume. 

This has consequences for the value of the apparent temperature measured
by an isotope ratio. Because the isotopic yield ratio depends exponentially 
on the temperature (${P_1\over P_2} \propto exp -(E_1-E_2)/T$) 
the mean value of the  
apparent temperature $T_{app} = (E_2-E_1)/\ln <{P_1\over P_2}>$, where
$<{P_1\over P_2}> = \int dT f(T){P_1\over P_2} / \int dT f(T)$, differs
from the mean value $<T>$ of the temperature distribution f(T)
in the subsystem of the fragments. On the other hand the apparent temperature is
bounded from above because above a critical temperature between 6 and 8 MeV,
depending on the microcanonical program, fragments are not
stable anymore \cite{grr}. Therefore the increase of the apparent
fragment temperature is smaller than that of $<T>$.
That may be the
origin of the observation that the mean value of the apparent temperatures
increases slowly with increasing excitation energy.

Consequently, the argument, that the experimental data present evidence
that in nuclear reactions a liquid gas phase transition can be
observed, which was advanced in ref. \cite{po}, has two shortcomings.
First, the temperature of the system extracted from the isotope ratios 
has large error bars, and hence it is not possible to distinguish between an 
(expected) increase of the temperature with beam energy and a constant value 
(which has been interpreted as sign of a latent heat). Second, the
nonlinearity of the isotope ratio with temperature in connection with upper
limit due to the instability of the fragments limits the apparent 
temperature to a small interval. Its dependence on the excitation energy of the
system is small as compared to that of the mean value of the
temperature distribution of the subsystem. 

\section{conclusions}
First of all we have found that the nuclear systems of the size as expected to
be formed in heavy ion reactions are too small to be susceptible to a 
canonical description but have to be analyzed in microcanonical approaches
because the results of both approaches differ substantially.

Therefore, the conjecture of Toeke et al. that the functional
dependence of the kinetic energies of fragments and light charged particles on
the observed number of IMF's may be used to distinguish between a statistical
and a dynamical reaction mechanism,
which was based on a canonical description, cannot be substantiated 
by detailed calculations.
On the contrary, dynamical and microcanonical calculations give almost identical
results. QMD calculations predict that the reaction
never comes to a statistical equilibrium and is completely determined by
the dynamics \cite{neb,reg}. Hence for a decision upon the reaction mechanism
one has to study other observables. 

Another result of the microcanonical analysis is the observation of 
large fluctuations of the
temperature if it is determined from fragments only. Consequently, due to the 
nonlinearity of the isotopic yield ratio as a function of the temperature and
due to the fact that at high temperature, fragments do not survive, the
apparent temperature measured from the isotopic yield increases slower as
compared to 
the mean value of the temperature distribution of the
fragments. In any case the relation between the apparent
temperature and the true temperature $1/T ={\partial S\over \partial E}$ is not trivial.
Hence the apparent temperature can not serve as a measure for the system 
temperature.
If the fluctuation of the temperature distribution are properly taken into
account, it is impossible to distinguish between a first order phase transition
(assuming it manifests itself in finite systems as a latent heat) and an 
increase of the apparent temperature with increasing beam or excitation energy.
Thus also the claim that data present evidence for a first order phase
transition is premature. These observations of our simple model remains valid
also in the framework of microcanonical models as can inferred from a comparison
of \cite{grr} and \cite{the}.

Acknowledgment: We thank R.Bougault et M. D'Agostino  for communicating to us 
the results of their SMM calculations which have been used for this analysis.



\begin{thebibliography}{99}
\bibitem{toek1} J. T\~oke et al.,Phys. Rev. Lett. {\bf{77}} 3514 (1996)
\bibitem{po} J. Pochodzalla et al.' Phys. Rev. Lett. {\bf 75} 1040 (1995)
\bibitem{gol} A.S. Goldhaber, Phys. Lett {\bf B53}  306 (1974)
\bibitem{hu} J. Aichelin, J. H\"ufner and R. Ibarra, Phys. Rev. C {\bf C30}, 
107 (1984).
\bibitem{hue} J. Huefner, Phys. Lett. {\bf B 173}, 373 (1986) 
\bibitem{fi}M.E. Fischer, Physics {\bf 3}, 255 (1967)\\
C.B. Chitwood et al. Phys. Lett. {\bf B131} , 2897 (1983)
\bibitem{bon95} J.P. Bondorf, A.S. Botvina, A.S. Iljinov, I.N. Mishustin, K.
Sneppen, Phys. Rep. {\bf{257}}, 133 (1995)
\bibitem{aic} J. Aichelin, Phys. Rep. {\bf 202}, 233 (1991), and references
therein.
\bibitem{neb} R. Nebauer et al. Nucl. Phys. {\bf A650}, 65 (1999)
\bibitem{qmdh} see www-subatech-in2p3.fr/~theo/qmd
\bibitem{reg} R. Nebauer et al. Nucl. Phys. {\bf A658}, 67 (1999)
\bibitem{gr} J. Bondorf et al. Phys. Rep. {\bf 257}, 133 (1995)
\bibitem{gs}D.H.E Gross and K. Sneppen, Nucl. Phys. {\bf A 567}, 317 (1993)
\bibitem{sal} S. Salou, Thesis, University of CAEN, France, GANIL T 97-06
\bibitem{mar} N. Marie, Thesis, University of CAEN, France, GANIL T 95-04 
\bibitem{bb} W. M\"uller et al., Phys. Lett. {\bf B 298}, 27 (1993) 
\bibitem{grr}D.H.E. Gross, K. Sneppen Nucl. Phys. {A567}, 317 (1994)
\bibitem{the} T. Odeh, Doctoral Thesis, Univ. Frankfurt/Germany (1999) 
fig. 61, available at the home page of Aladin/GSI on the www.






\end{thebibliography}
\end{document}